\documentclass[preprints,article,accept,pdftex,moreauthors]{Definitions/mdpi}

\firstpage{1} 
\pubvolume{1}
\issuenum{1}
\articlenumber{0}
\pubyear{2023}
\copyrightyear{2023}
\datereceived{ } 
\daterevised{ } 
\dateaccepted{ } 
\datepublished{ } 
\hreflink{https://doi.org/} 

\Title{Observing Dusty Star-Forming Galaxies at the Cosmic Noon through Gravitational Lensing: Perspectives from New Generation Telescopes}

\TitleCitation{Observations of Strongly Lensed DSFGs}

\Author{Marika Giulietti $^{1}$*\orcidA{}, Giovanni Gandolfi $^{4}$\orcidB{}, Marcella Massardi\orcidC{}$^{2, 3}$, Meriem Behiri $^{1,2}$\orcidD{} and Andrea Lapi $^{2}$\orcidE{}}

\AuthorNames{Marika Giulietti, Giovanni Gandolfi, Marcella Massardi, Meriem Behiri, and Andrea Lapi}

\AuthorCitation{Giulietti M.; Gandolfi G.; Massardi M.; Behiri M.; Lapi A.}

\address{%
$^{1}$ \quad INAF - Istituto di Radioastronomia,
Via Gobetti 101, 40129 Bologna, Italy\\
$^{2}$ \quad Scuola Internazionale Superiore di Studi Avanzati, Via Bonomea 265, 34136 Trieste, Italy \\
$^{3}$ \quad INAF - Istituto di Radioastronomia - Italian ALMA Regional Centre, Via Gobetti 101, 40129 Bologna, Italy \\
$^{4}$ \quad Dipartimento di Fisica e Astronomia, Univ. di Padova, Vicolo dell’Osservatorio, 3, 35122 Padova, Italy
}

\corres{Correspondence: m.giulietti@ira.inaf.it}

\abstract{Gravitational lensing, a compelling physical phenomenon, offers a unique avenue to investigate the morphology and physical properties of distant and faint celestial objects. This paper seeks to provide a comprehensive overview of the current state of observations concerning strongly lensed Dusty Star-Forming Galaxies. Emphasis is placed on the pivotal role played by cutting-edge facilities like the James Webb Space Telescope and the Square Kilometer Array Observatory. These advanced instruments, operating at disparate ends of the electromagnetic spectrum, in conjunction with the amplifying effect of gravitational lensing, promise significant steps in our understanding of these sources. The synergy between these observatories operating at two opposite ends of the spectrum, is poised to unlock crucial insights into the evolutionary path of high-redshift, dust-obscured systems and unravel the intricate interplay between Active Galactic Nuclei and their host galaxies.}


\begin{document}




\section{Introduction}

Dusty Star-Forming Galaxies (DSFGs; see e.g. \cite{Casey2014} for a review) provide a unique window into the processes driving intense star formation, galaxy interactions, and feedback mechanisms in the early Universe. By studying these extreme objects, one can gain insights into the processes that shape galaxies over billions of years, shedding light on the overall history of galaxy evolution.

During the so-called \textit{cosmic noon}, galaxies exhibited intense star formation within dust-rich environments, arising from rapid gas collapse within central regions of dark matter halos. This vigorous star formation phase enriched inner regions of the galaxy with dust (via supernovae explosions) and facilitated accelerated feeding of central black holes.

This high activity led to the absorption of energetic UV photons by newborn stars, heating dust to temperatures of 20-80 K, and subsequent re-emission in FIR/sub-mm wavelengths.
Instruments such as the Submillimetre Common-User Bolometer Array (SCUBA; \cite{Holland1999MNRAS.303..659H}) equipped on the James Clerk Maxwell Telescope, (JCMT\footnote{\url{https://www.eaobservatory.org/jcmt/}})  or the Max Planck Millimeter Bolometer (MAMBO\footnote{\url{https://astro.uni-bonn.de/~bertoldi/projects/mambo/}}; \cite{Kreysa1998SPIE.3357..319K}) revealed that these galaxies are extremely bright in sub-millimetre (sub-mm) wavelengths (at $\lambda=850\, \mu$m) and for this reason labelled Sub-millimetre Galaxies (SMGs; \cite{Smail1997}).

DSFGs feature extreme properties, such as incredibly high star formation rates (100-1000 M$_{\odot}$ yr$^{-1}$), clumpy morphologies, and high mass content ($M_{\star} \gtrsim 10^{10}\, {\rm M}_{\odot}$).

These extreme objects are found to be $\sim1000$ times more abundant at high redshifts (e.g. \cite{Blain2002}) with respect to the nearby Universe, dominating the cosmic star formation between redshifts $z \simeq 2-4$ (e.g. \cite{Gruppioni2013}). At higher redshifts, the volume density of DSFGs is still unconstrained (e.g. \cite{Casey2018ApJ...862...77C}), given that studies above $z\gtrsim4$ are mainly based on UV-selected samples such as Lyman-break galaxies (LBGs). On the one hand, observations point towards a rapid decline of the SFR density towards higher redshifts (\cite{Bouwens2015}; \cite{McLeod2016}; \cite{Ishigaki2018}). On the other hand, long-wavelength observations seem to point toward a flatter decline of the SFR density at z$\gtrsim3$ (\cite{Gruppioni2013}; \cite{Rowan-Robinson2016}; \cite{Gruppioni2020}; \cite{Talia2021}; \cite{Enia2022}), hinting towards a possible major role of dust-rich galaxies in the stellar mass assembly.

The peak in the number density of DSFGs at $z\sim 2$ (\cite{Casey2014}) mirrors the peaks in both the Star Formation Rate Density (SFRD) and Black Hole Accretion History (\cite{Madau2014}), suggesting that DSFGs have made a substantial contribution to the Universe's SFRD and overall stellar mass density and can serve as ideal laboratories to test the AGN-galaxy co-evolution.
Observations (e.g. \cite{Harikane2023}; \cite{Juod2023MNRAS.525.1353J}; \cite{Yang2023}) are indeed pointing towards the presence of AGNs up to the epoch of reionization (EoR), indicating their presence and activity in the early universe and their impact on the evolution of the host galaxy.
In fact, the presence of tight relations between the mass of the central SMBH and properties of the host galaxies such as the velocity dispersion, the stellar mass, the luminosity and morphology (e.g. \cite{Kormendy1995}; \cite{Magorrian1998AJ....115.2285M}; \cite{Ferrarese2000, Ferrarese2005}; \cite{McLure2004}; \cite{Kormendy2009, Kormendy2013}; \cite{Shankar2016}), strongly hints towards the presence of a parallel evolution.
Furthermore, another hint comes from the SFRD trend, which mirrors the one of the BH accretion rate density (\cite{Boyle1998}; \cite{Franceschini1999}; \cite{Delvecchio2014}; \cite{Madau2014}; \cite{Aird2015}; \cite{Mancuso2016a}).


Observations strongly imply that DSFGs underwent a transition to become Early-Type Galaxies (ETGs) observed in the local Universe. This transformation involved shifting from an active phase of star formation to a more passive state, where the stellar and AGN feedback may have played a significant role (see e.g. \cite{Peng2010}; \cite{Cimatti2008}; \cite{Behroozi2013}; \cite{Simpson2014}; \cite{Toft2014}; \cite{Aversa2015}; \cite{Mancuso2016a}; \cite{Oteo2017}; \cite{Scoville2017}).

The significance of DSFGs therefore extends to several key aspects of cosmic evolution. They play a central role in the evolution of AGN, the formation of galactic spheroids, and the development of massive ETGs. By serving as a bridge between actively star-forming and passive galaxies, DSFGs have a pivotal role in shaping our understanding of early cosmic star formation.

Nonetheless, understanding the observed properties of these galaxies and unravelling the mechanisms governing the co-evolution of galaxies and black holes remains a topic of ongoing discussion.

Several theoretical scenarios emerged to face these challenges. 
The \textit{major merger}-induced starburst envisages the merging of gas-rich spirals at high redshift as the main route toward building up massive ellipticals and triggering their star formation and BH activity (e.g. \cite{Bower2006}; \cite{Croton2006}; \cite{Hopkins2006}; \cite{Benson2010}, \cite{Fanidakis2012}; \cite{Somerville2015}).
Other frameworks assume that star formation and BH accretion are driven by steady cold gas streams along filaments of the cosmic web (\cite{Dekel2009}; \cite{Bournaud2011}), or that they occur \textit{in situ} and are mainly governed by self-regulated baryonic physics, in particular by energy feedback from SNe and the central nucleus (e.g., \cite{Granato2004}; \cite{Lapi2006, Lapi2011}; \cite{Aversa2015}; \cite{Mancuso2016a}).
In this latter view, the merging events play a minor role in the late slow stage of the evolution of the quenched galaxy and are mainly involved in the mass increase to the values observed in the local Universe (\cite{Lapi2018}).

Over the last twenty years, observations have provided insights into the nature of DSFGs, uncovering their statistical features as a result of larger numbers being identified through wide-area surveys ($>0.1-100 \rm deg^2$; e.g. \cite{Smail1997}; \cite{Bakx2018}; \cite{Scott2008}; \cite{Eales2010}; \cite{Aretxaga2011}; \cite{Casey2012b}; \cite{Casey2012c}) conducted with both ground-based (e.g. JCMT), and space-based facilities (e.g. the \textit{Spitzer\footnote{\url{https://www.spitzer.caltech.edu/}}} and \textit{Herschel\footnote{\url{https://www.herschel.caltech.edu/}}} Space Observatories). 

However, a comprehensive understanding of these galaxies and their evolution goes beyond evaluating their integrated properties retrieved only in sub-mm bands, as vital evolutionary processes occur on sub-galactic scales, encompassing stellar and AGN feedback, star formation triggers and quenching, spheroid formation, and galaxy-black hole co-evolution, requiring a variety of multi-wavelength data.
Indeed, what has become evident in the recent decades of investigations, is the necessity of gaining a collective view of the different emission properties of galaxies, racking the evolution of their luminosity across cosmic time from far-UV (FUV) to radio wavelengths (e.g. \cite{Smolcic2017}; \cite{Elbaz2018}).

Specifically, the light coming from short-lived massive stars is predominantly traced by the rest-frame UV-continuum emission, which can be a direct measure of the instantaneous SFR density. 
The bulk of the galaxy's stellar mass is constituted by evolved near-solar massive stars, whose light can be traced in the rest-frame NIR emission. Furthermore, UV emission is absorbed by the interstellar dust and re-emitted as thermal infrared radiation, rendering the FIR emission from dusty starburst galaxies a sensitive tracer of young stellar populations and SFRD.
Under the dust blanket, X-ray DSFGs
observations revealed on-growing central SMBHs, preluding to the future phase in which
they will quench star formation and evacuate gas and dust from the host galaxy (e.g. \cite{Rodighiero2015}).
The AGN’s emission heats up the surrounding dust creating an excess in the mid-infrared
(MIR) band (if compared with the MIR emission of galaxies with intense star formation but no strong AGN contribution). Such MIR excess is an important hint signalling the possible presence of AGN activity in these galaxies (\cite{Mullaney2011MNRAS.414.1082M}; \cite{Stern2012}; \cite{Bonzini2013}, \cite{Padovani2015}).
At longer wavelengths, radio emission serves as another significant indicator of star formation unaffected by dust obscuration. Its primary component emerges from massive stars that explode as Type II supernovae, primarily radiating their energy in the radio band through synchrotron emission at the end of their life. Additionally, the radio emission contains a secondary component, originating from the free-free emission of the hot and ionised HII regions. An excess in the radio band can also be indicative of AGN activity (\cite{Bonzini2015}; \cite{Delvecchio2017A&A...602A...3D}).

For all these reasons, multi-wavelength observations and modelling of emissions from galaxies' components are nowadays considered the most promising approach to address the open questions in DSFGs' description across cosmic time (see \cite{Tacconi2020} for a review).

Nonetheless, grasping the intricate spatial distribution and dynamics of individual molecular clumps, often on sub-kiloparsec scales, remains challenging. 

Early studies of DSFGs were hindered by the limited angular resolution of (sub-)mm instruments, which could resolve structures only on scales of a few kiloparsecs. Moreover, such observations typically required substantial investments of observing time (\cite{Tacconi2013ApJ...768...74T}; \cite{Hodge2015ApJ...798L..18H}).



The advent of the Atacama Large Millimeter/submillimeter Array (ALMA\footnote{\url{https://www.almaobservatory.org}}) in recent years has significantly advanced our understanding (see e.g. \cite{Hodge2020}). For instance, ALMA continuum observations provided insights into the properties of the bulk of dust content (composed by the coldest dust grains) by tracing the Rayleigh-Jeans tail of thermal dust emission of DSFGs at $z \sim 2-3$.

In this sense, high-resolution ALMA observations, with sub-arcsecond spatial resolution, have been of great impact. High-z DSFGs found both above and on the main sequence, have been found to host a compact ($\sim 1 - 5$ kpc, \cite{Simpson2015}; \cite{Barro2016a}; \cite{Oteo2016}; \cite{Barro2017}, \cite{Fujimoto2017}; \cite{Tadaki2017a}; \cite{Tadaki2017b} \cite{Talia2018}; \cite{Nelson2019}; \cite{Pantoni2021}) and central concentrated dust continuum. The sizes of this compact dust component, are often found to be consistent with the sizes inferred from radio emission, however, they differ from optical measurements ($\sim 2 - 10$ kpc; \cite{Rujopakarn2016}; \cite{Nelson2019}; \cite{Pantoni2021}). Yet a fraction ($\sim 20\%$) of sub-mm-selected DSFGs remains undetected in deep optical/near-IR (NIR) images, such as the one from the Hubble Space Telescope (HST\footnote{\url{https://hubblesite.org/home}}; e.g. \cite{Franco2018}; \cite{Wang_T_2019}; \cite{Gruppioni2020}). 

Further complexities arise from the multi-wavelength morphology of these galaxies. Recent studies employing ALMA and HST imaging have highlighted clumps in the optical rest-frame and smoother FIR rest-frame emissions with compact cores and extended disks (e.g. \cite{Rujopakarn2019}; \cite{Rujopakarn2023}). Clumpy structures have also been observed with ALMA in the FIR rest-frame continuum (\cite{Hodge2019}), while larger scales ($\sim$ few kpcs) reveal isolated and disturbed morphologies (\cite{Hodge2016}; \cite{Elbaz2018}), often requiring multiple-component models for multi-wavelength emission.
Spatially-resolved radio emission provides insights to distinguish between the stellar emission from AGN contributions, though such studies are limited due to extensive integration times required for high-redshift analysis. 


Despite progress, a complete physical understanding of high-z DSFGs remains elusive. A comprehensive approach involving galaxy properties characterisation, molecular spectral lines, and spatially-resolved multi-band imaging is crucial. This approach provides insights into mechanisms and processes across evolutionary stages, offering clarity on the role of DSFGs within the broader context of galaxy formation and evolution.

Strong gravitational lensing, which occurs when the gravity of massive objects such as galaxies or galaxy clusters bends and distorts light from background sources, offers an exceptional opportunity for astronomers to investigate the properties of these objects. This distortion results in the formation of multiple images, such as arcs or rings, providing valuable insights into both the distribution of mass in the universe (\cite{Grillo2008}; \cite{Oguri2012}; \cite{Eales2015}), including the nature of dark matter (e.g. \cite{Vegetti2009}; \cite{Hezaveh2016}), as well as the properties of both the lensing object and the background source.

There are two primary benefits when studying strongly lensed galaxies at high redshift. Firstly, the source's surface brightness is conserved while its apparent luminosity is linearly magnified with the magnification factor $\mu$. This unique combination allows the observations of regions in the luminosity-redshift space of faint astrophysical sources that would otherwise remain inaccessible.
Next, gravitational lensing distorts the shapes of the sources. For instance, when an event is produced by a lensing galaxy, it leads to the observation of multiple images of the object, which appear elongated or "stretched" to the observer by a factor $\mu^{1/2}$. As a result, the angular sizes of the source are amplified, providing an enhanced view and enabling a more detailed examination of the object's features. 

Gravitational lensing has been observed to impact the apparent luminosity of distant populations of objects which would otherwise remain undetected, such as Lyman-Break Galaxies (LBGs), with a higher fraction of magnified objects occurring at greater redshifts. For instance, approximately $10\%$ of bright LBGs at a redshift around $z\sim 7-8$ have been identified as lensed (\cite{Barone-Nugent2015MNRAS.450.1224B}).

These advantages are particularly powerful when studying DSFGs, for which (sub-)mm observations are often limited by poor angular resolution.
For example, by exploiting the extreme angular resolution of ALMA, it has been possible to resolve dust continuum and the kinematic of molecular gas down to a $\sim$10-100 pc-scale (e.g. \cite{Partnership2015}; \cite{Rybak2015a, Rybak2015}), comparable to low-redshift objects (\cite{Swinbank2010}).

Strong gravitational lensing, therefore, offers a powerful way to zoom in on the star formation processes and interstellar medium of dusty galaxies at the Cosmic Noon.


In this paper, we aim to summarize the current status of observations of strongly lensed DSFGs, highlighting the role of new and upcoming facilities such as the James Webb Space Telescope (JWST\footnote{\url{https://webbtelescope.org/home}}) and the Square Kilometer Array Observatory (SKAO\footnote{\url{https://www.skao.int/}}) in facilitating a significant advancement in the exploration of these objects.
The plan of the paper is the following: in Section \ref{sec:1} we describe the main surveys and selection methods employed for identifying strongly lensed systems bright in the (sub-) millimetre range. In Section \ref{sec:2} we outline the main observational strategies which can be employed to investigate strongly lensed DSFGs from observations. Section \ref{sec:3} provides future perspectives and opportunities offered by upcoming radio band facilities and their precursors, while Section \ref{sec:4} focuses on the new window in the Near- and Mid-Infrared regime opened by JWST. Finally, a summary of the key points of the paper is given in Section \ref{sec:5}.

\section{Selecting Strongly Lensed DSFGs in the Sub-Millimeter Regime}\label{sec:1}

Models predicting the number counts of gravitationally lensed sources in the mm and sub-mm waveband
(\cite{Blain1996}; \cite{Perrotta2002}; \cite{Lapi2006}; \cite{Negrello2007}; \cite{Negrello2010}; \cite{Lapi2011}) envisage a steep luminosity function for SMGs, resulting in a sharp drop in the number counts at the brightest fluxes ($\simeq 80-100$ mJy at 500 $\mu$m). Beyond this threshold, various galaxy populations can be detected, such as low-redshift (z$\lesssim$0.1), late-type galaxies, flat spectrum radio sources, Hyper Luminous Infrared Galaxies (HyLIRGs), and gravitationally lensed DSFGs at z$\gtrsim 1$. By adopting this simple flux density threshold, combined with shallow optical and radio data to identify and remove contaminants, gravitationally lensed systems can be efficiently selected.
Given the low predicted surface density of these systems ($\lesssim 0.5$ deg$^{-2}$), large area surveys are required to detect a significant number of lensing events.

\begin{figure}
    \centering
    \includegraphics[width=0.8\textwidth]{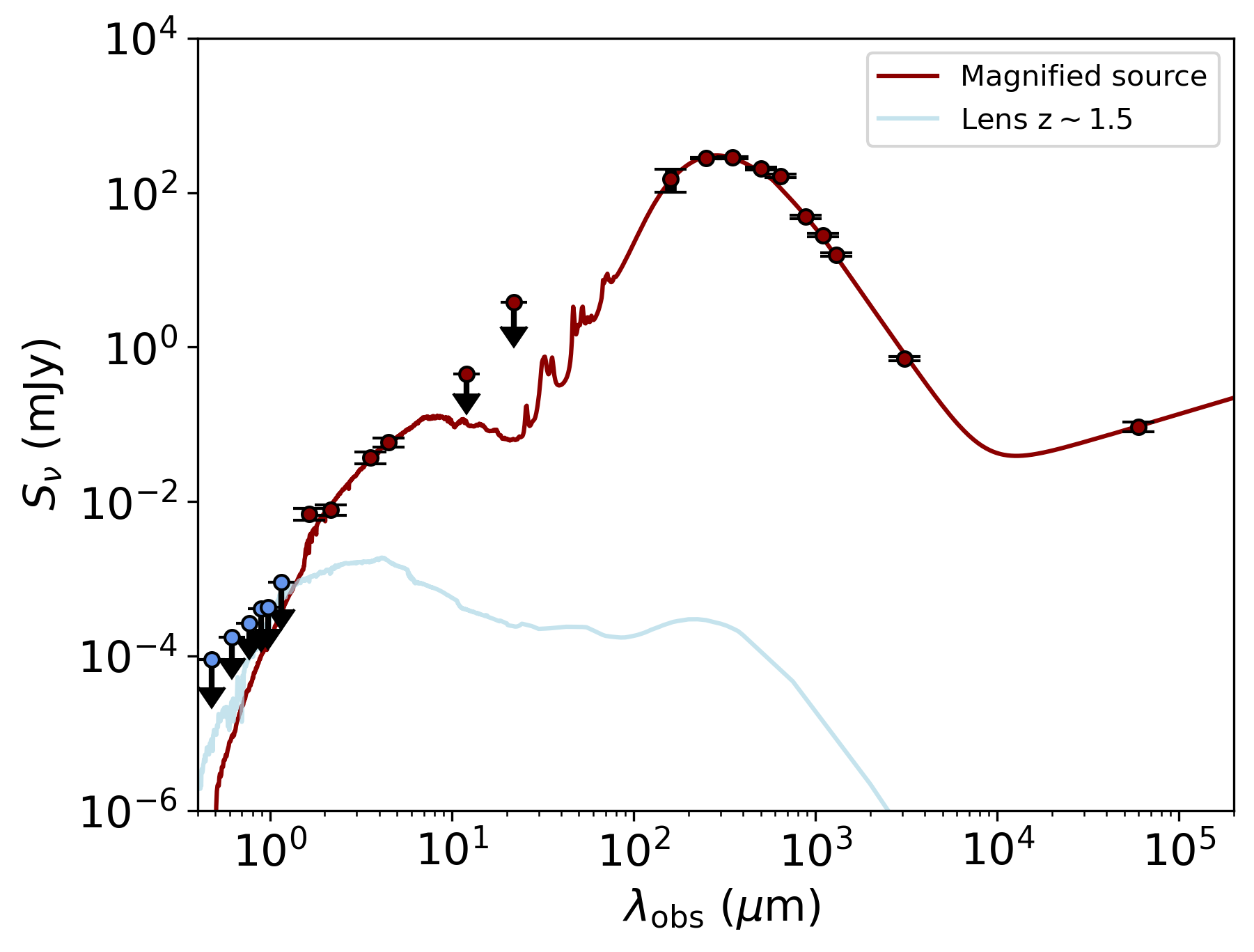}
    \caption{Example of the SED of a strongly lensed system, namely G12v2.43, analysed in \cite{Giulietti2023} and \cite{Perrotta2023}. Photometric points from \cite{Giulietti2023} are shown with circles, red and blue lines are respectively the best-fit SED of the (magnified) background source and the lens (assumed to be a passive galaxy at $z\gtrsim 1.5$). In this case, the foreground object remains undetected in the optical/NIR bands and its contribution to this regime is negligible.}
    \label{fig:1}
\end{figure}

In FIR/sub-mm bands, high-z strongly lensed DSFGs are exceptionally bright and the interference stemming from the foreground lens – typically a passive elliptical galaxy at $z\lesssim 1$ (\cite{Negrello2014}) – is minimal (see Figure \ref{fig:1}). Consequently, lens modelling and the source morphology reconstruction remain largely unaffected by uncertainties associated with lens subtraction, an obstacle often encountered in optically selected lensing systems.

Furthermore, this selection method exclusively harnesses the attributes of the background galaxy itself, specifically its sub-mm flux density. As a result, any potential bias against the lens's redshift and mass is circumvented. 


In this sense, the capabilities of the \textit{Herschel} Space Observatory and the comprehensive \textit{Planck} all-sky survey at sub-mm wavelengths (\cite{Pilbratt2010}; \cite{Canameras2015}), along with the achievements of the South Pole Telescope (SPT\footnote{\url{https://pole.uchicago.edu/public/Home.html}}, \cite{Vieira2010}; \cite{Carlstrom2011}) and the Atacama Cosmology Telescope (ACT\footnote{\url{https://act.princeton.edu/}}, \cite{Swetz2011}; \cite{Marsden2014}; \cite{Rivera2019}) in the mm frequency bands has been amply proved. 

For what concerns \textit{Herschel}-selected samples, 11 lensed galaxies have been identified over the 95 deg$^2$ of the \textit{Herschel} Multi-tired Extragalactic Survey (HerMES, \cite{Oliver2012}; \cite{Wardlow2013}); other 77 candidate lensed galaxies were found in the \textit{Herschel} Large Mode Survey (HeLMS, \cite{Oliver2012}; \cite{Nayyeri2016}), and in the \textit{Herschel} Stripe 82 Survey (HerS, \cite{Viero2014}). 
Two samples of strongly lensed DSFGs were selected in the \textit{Herschel}- Astrophysical Terahertz Large Area Survey (H-ATLAS, \cite{Eales2010}), which is the largest area ($\sim 600$ deg$^2$) extragalactic survey undertaken by \textit{Herschel}. 
80 candidate strongly lensed DSFGs were selected adopting the photometric criteria: $S_{500\mu\rm m} \gtrsim 100 \rm mJy$ (\cite{Negrello2017}), and, with the recent Third Data Release 11 more candidates in the Southern Galactic Pole of the survey were found by \cite{Ward2022}. 
The selection criteria adopted by \cite{Negrello2017} has been extended by \cite{Bakx2018} by including DSFGs with (sub-)mm color-estimated photometric redshift z$_{\rm phot}>2$, and $S_{500\mu m} > 80$ mJy. This sample, namely the Herschel Bright Sources (HerBS) sample, is expected to contain both lensed and unlensed sources, given that for fluxes $S_{500\mu m} < 100$ mJy the number density of unlensed DSFGs increases exponentially. 

In the sub-mm regime, also the Planck\footnote{\url{https://www.cosmos.esa.int/web/planck}} All-Sky Survey to Analyze Gravitationally lensed Extreme Starbursts (PASSAGES; \cite{Harrington2016MNRAS.458.4383H}; \cite{Berman2022}) led to a list of 30 candidate lensed high-redshift ($1 \lesssim z \lesssim 3$) galaxies with $S_{350\mu \rm m}\gtrsim 100$ mJy cross-matched with \textit{Herschel} and the Wide-field Infrared Survey Explorer (WISE\footnote{\url{https://wise2.ipac.caltech.edu/docs/release/allsky/}}). Recently, the multi-wavelength lens modelling and analysis of 15 out these 30 sources was presented in \cite{Kamieneski2024}, combining sub-arcsecond observations from HST, the Gemini Telescope\footnote{\url{https://www.gemini.edu/}}, ALMA and the Very Large Array (VLA\footnote{\url{https://public.nrao.edu/telescopes/vla/}}).

At longer wavelengths, the SPT-Sunyaev–Zel'dovich survey (\cite{Mocanu2013}) identified 81 strongly lensed DSFGs over an area of 2500 deg$^2$ by adopting a slightly different selection method (S$_{\rm 1.4 mm}>20$ mJy, \cite{Vieira2013}; \cite{Weiss2013}; \cite{Spilker2016}; \cite{Everett2020}; \cite{Reuter2020}).
Because of the longer selection wavelength, SPT and ACT strongly lensed sources are found to be brighter in the FIR and at higher redshift ($z\gtrsim 6$) with respect to \textit{Herschel} selected samples (\cite{Weiss2013}; \cite{Bethermin2015}; \cite{Su2017MNRAS.464..968S}).

\section{Studying lensed DSFGs from observations}\label{sec:2}

Sub-mm-selected samples of lensed galaxies are ideal for follow-up observations aimed at understanding the detailed properties of starburst phenomena in distant galaxies.
These properties can be investigated through two primary and complementary strategies, harnessing the potential of multi-wavelength data from strongly lensed high-redshift DSFGs.

\subsection{Studies of individual sources}\label{sec:individual_sources_studies}

The second method centres on the study of limited samples or individual objects that possess an extensive collection of high-quality photometric and spectroscopic data (e.g. \cite{Partnership2015}; \cite{Dye2015,Dye2022}; \cite{Giulietti2023}; \cite{Massardi2017}; \cite{Perrotta2023}; \cite{Rybak2015}; \cite{Kamieneski2023}; \cite{Borsato2023} and references therein). This approach aims to achieve a profound understanding of the ongoing astrophysical processes inside the galaxy.

Up to the advent of JWST, achieving high-resolution imaging has been feasible in both the optical/NIR domain, especially thanks to HST or adaptive optics systems such as the \textit{Keck} Observatory and in the mm/radio spectrum using ground-based interferometers such as ALMA and the VLA.
The combination of high-resolution imaging in both the long and short-wavelength regimes is particularly important in studying a lensing system. Indeed, one of the significant uncertainties relies on the determination of the stellar component of the background DSFG, whose light can be captured in the observed optical/NIR regime. However, the light of the lensing system at these wavelengths is intrinsically faint because of the dust obscuration and the redshift and is often dominated by the foreground lens' emission (\cite{Negrello2014}).
To overcome this issue, accurate modelling and subtraction of the lens light profile are required to detect the faint unobscured stellar emission from the background source.
This is necessary to perform accurate Spectral Energy Distribution (SED)-fitting and disentangling the contributions from galaxies in the lensing system, and can only be achievable with high-quality imaging.
Conversely, achieving high-resolution follow-up observations in the sub-mm and mm wavelengths is essential for unveiling the gas and dust components of the background galaxy (\cite{Negrello2014}, \cite{Borsato2023}).
Additionally, due to the brightness of DSFGs in this range, enhanced by the negative k-correction and the minimal foreground lens contamination, continuum observations at high angular resolution prove to be highly effective in the visual identification of lensed objects. These observations, in particular those conducted with ALMA and reaching extreme angular resolutions and sensitivities (e.g. \cite{Dye2014}; \cite{Rybak2015}; \cite{Spilker2016}; \cite{Massardi2017}; \cite{Rizzo2021}), reveal distinct arcs and multiple images of the source. This requires less effort compared to optical/NIR observations where the subtraction of the lens' light is needed. This becomes particularly crucial because, despite magnification effects, the combination of dust obscuration and high redshift can hinder the detection of the unobscured stellar component in the optical/NIR for these sources (e.g., \cite{Giulietti2023}).

Multi-band observations can also be combined in order to improve the lens modelling and ease the identification of image families of the lensed source (see \cite{Kamieneski2024}).

Spectroscopic data are extremely important to confirm the source's redshift, kinematic, and gas content (e.g. \cite{Tadaki2015}, \cite{Barro2016a,Barro2016b}, \cite{Decarli2016}, \cite{Talia2018}, \cite{Rizzo2020}, \cite{Pantoni2021}, \cite{Rizzo2021, Rizzo2022}). Observing spectral lines in high-z DSFGs requires an enormous amount of time (several hours), even with ALMA, and therefore it is often only feasible for individual sources. 
The magnification offered by strong lensing improves the performance of these observations.
For example, the redshifts of 13 bright galaxies identified in H-ATLAS with S$_{500\mu m}\gtrsim 80$ mJy were determined in\cite{Neri2020}, including several lensing systems. They established robust spectroscopic redshifts for 12 individual sources, relying on the identification of at least two emission lines. Building on the success of this initial study, a comprehensive and extensive survey, known as z-GAL (\cite{Cox2023}), has recently been conducted employing the Northern Extended Millimeter Array (NOEMA\footnote{\url{https://iram-institute.org/science-portal/noema/}}). This survey yielded dependable redshifts for all 126 bright \textit{Herschel}-selected SMGs with 500 $\mu$m fluxes exceeding 80 mJy. These sources were chosen from the H-ATLAS and HerMES fields situated in the Northern and equatorial regions of the sky. 

\noindent
The BEARS survey is another recent redshift campaign (\cite{Urquhart2022}, \cite{Bendo2023}, \cite{Hagimoto2023}) performed with ALMA using both the 12m and the Atacama Compact Array (ACA) arrays and targeting the brightest sources in the southern field of the H-ATLAS survey. While 81 sources were targeted, 142 objects were resolved and 71 were spectroscopically confirmed through the detection of [CI], CO, and H$_2$O lines.
The outcomes of this redshift campaign open up several follow-up observations. \cite{Bendo2023} measured the continuum emission reconstructing the SED of these data, and \cite{Hagimoto2023} estimated the physical conditions of the ISM through the detected spectral lines [CI], CO, and H$_2$O lines.

\noindent
Spectroscopic observations of strongly lensed sources can also be exploited to infer important dynamical properties of distant galaxies with unprecedented angular resolution. For instance, the dynamics of the strongly gravitational lensed galaxy SPT-S J041839-4751.9 at $z=4.2$ has been investigated in \cite{Rizzo2020}. By applying a three-dimensional lens-kinematic modelling technique to ALMA data, authors found evidence of a dynamically cold and highly star-forming disc. 

\noindent
This approach has been extended in \cite{Rizzo2021} for a sample of five strongly lensed DSFGs, exploiting ALMA spectroscopic [CII] data, and reconstructing their kinematics down to a scale of $\sim 200$ pc.

\subsection{Studies of statistically significant samples}

The first approach involves the analysis of statistically relevant samples of strongly lensed DSFGs. 
This method is centred on the selection of a group of objects sharing common observational traits. The robustness of this method lies in its ability to yield a coherent and representative so to be statistically significant sample containing a substantial number of sources, which is essential for analysing physical attributes. However, a notable drawback is the scarcity of data, particularly in specific regions of the electromagnetic spectrum. This scarcity can introduce uncertainties in photometric redshifts, and in some cases, necessitate arbitrary assumptions regarding under-sampled galaxy properties. 



\noindent
In the context of gravitational lensing, this approach benefits from the inherent magnification in the flux density of the objects in question. Consequently, it offers the great advantage of broadening studies to encompass galaxies that possess intrinsically lower flux densities and often fall below the confusion limit (e.g. \cite{Stacey2018, Stacey2019}; \cite{Giulietti2022}; \cite{Gururajan2023}). 

Gravitational lensing also offers an extra advantage due to its unique systematic biases when contrasted with field sources. While observations in the field tend to be biased towards sources with high luminosity or low redshifts, gravitationally lensed sources are biased towards compact, higher redshift objects (typically $z > 1$). Moreover, they are less biased against high intrinsic luminosities (e.g. \cite{Swinbank2010}).

However, the primary drawback when examining extensive samples of lensed DSFGs arises from the lack of uniformity in the angular resolution of the data. This aspect becomes particularly critical in the determination of the intrinsic (de-magnified) physical properties. Lens modelling is indeed a process that can be accomplished solely for objects with resolved spatial features (\cite{Negrello2014}). Consequently, when lens modelling is not accessible for all objects within the sample, one must resort to making assumptions about the value of the magnification factor (\cite{Gururajan2023}).

In the next Section we will focus on the physical information that the radio and Near/Mid-IR bands can reveal about strongly lensed DSFGs. These wavelength regimes are relevant to new-generation instruments such as JWST or the upcoming SKAO and its precursors.

\section{Improving the Information in the Radio Continuum Domain}\label{sec:3}

In Section \ref{sec:1} we introduced the relevant role which DSFGs play in co-evolutionary models.
Addressing the topic of the co-evolution between galaxies and SMBHs involves not only a knowledge of the SFR but also of the AGN feedback mechanisms (jets or winds), concurring in the quenching of the star formation. Obtaining independent measurements of star formation and AGN activity is essential, especially in quasar-host galaxies which are rich in gas and dust (\cite{Hopkins2005}). 
One important issue regards the main driver of radio emission in heavily obscured accreting SMBHs (\cite{Alexander2012}), and its connection with galaxy properties such as the age, the sSFR, and the obscuration. 

From the Far-IR/Radio Correlation an excess in radio emission is expected from radio-loud AGNs, which is mainly detected at fluxes $S_{\rm 1.4\, GHz} \gtrsim 0.1\,$mJy (e.g. \cite{Mignano2008}). This population progressively recedes towards fluxes below this threshold, making way for the emerging category known as radio-quiet AGNs (e.g., \cite{Simpson2006}, \cite{Seymour2008}, \cite{Smolcic2008}). 
This also corresponds to a gradual change in the physical processes probed by deep radio surveys. However, the results of the radio emission mechanisms (AGN, star formation, or a composite of both) have been largely debated (e.g. \cite{Barthel2006}; \cite{Kimball2011}; \cite{Padovani2011}; \cite{Bonzini2013}; \cite{Condon2013}; \cite{Bonzini2015}; \cite{Padovani2015}; \cite{White2015}; \cite{Herrera-Ruiz2016}; \cite{Kellermann2016}; \cite{Herrera-Ruiz2017}; \cite{White2017}).

In \cite{Giulietti2022} authors show that the in-situ scenario provides a plausible evolutionary framework for the evolution of radio-loud AGNs (see also \cite{Mancuso2017}).
In this view, samples of strongly lensed quasars and DSFGs can be exploited to overcome the limits in sensitivity of the current surveys, where the majority of these objects often remain undetected. However, most lensed quasars still lie below the detection limits of all-sky surveys such as the Faint Images of the Radio Sky at Twenty-Centimeters (FIRST; \cite{Becker1995}) survey and even forthcoming surveys such as VLA Sky Survey (VLASS, \cite{Lacy2020}), and samples of strongly lensed objects are still limited.
Even though significant progress has been made in characterising the faint radio population with deep surveys of single fields, such as the VLA-COSMOS 3 GHz survey (\cite{Smolcic2017}), these predictions need to be tested in detail in order to fully understand the mechanisms that give rise to radio emissions in radio-quiet AGNs, in particular, the impact of radiative feedback processes and whether or not radio emission in these objects is primarily governed by star formation.
To tackle this issue, it is essential to examine the SED of a sufficiently large sample of galaxies across a broad spectrum of frequencies. This spectrum should extend from radio waves to the FIR range, with special attention given to encompassing the transitional zone between synchrotron/free emission in the radio and dust emission in the (sub-)mm domain.
In the future, the radio domain will soon open a new observational window, provided by the upcoming ultradeep radio continuum surveys planned on the SKAO and its pathfinder telescopes, such as the Australian Square Kilometre Array Pathfinder (ASKAP\footnote{\url{https://www.atnf.csiro.au/projects/askap/index.html}}), MeerKAT\footnote{\url{https://www.sarao.ac.za/science/meerkat/}}, and the LOw Frequency Array (LOFAR\footnote{\url{https://www.astron.nl/telescopes/lofar/}}; see \cite{Norris2013}, \cite{Prandoni2015}).

LOFAR opened the door to multifrequency (down to 60-50 MHz) deep radio surveys that have been recently conducted in the LOFAR Two-metre Sky Survey Deep Fields (LoTSS, e.g. \cite{Bonato2021}), covering wide-area fields
such as the Lockman Hole, reaching $\sim 6.6$ deg$^2$ (\cite{Prandoni2018}). 
A broad coverage of the radio spectra offers a means to refine the understanding of the source of radio emissions in observed objects and its correlation with the host galaxy's bolometric emission.

Recent studies have exploited the low-frequency regime ($\nu \lesssim 150$ MHz), currently explored by LOFAR (\cite{vanHaarlem2013}) and soon with SKAO, to detect the emission originated by star formation at high-z through the steep negative spectrum of synchrotron radiation, where the observed radio luminosities are higher. This alternative method to identify extreme cases in which the radiative output from the AGN is heating the cold dust at FIR wavelengths can also be applied in the case of gravitationally lensed systems, where the magnification in sizes and luminosity enable the study of the FIR and radio properties for individual objects otherwise undetected (see \cite{Stacey2019}).

SKAO path-finders such as ASKAP are built to perform high-speed surveys with an instantaneous field of view up to 30 deg$^2$.
The ASKAP Evolutionary Map of the Universe survey (EMU, \cite{Norris2011, Norris2017}) is planned to conduct all-sky deep observations. One of the key objectives of the survey is tracing the co-moving SFRD up to $z\sim2$ for starburst galaxies and up to $z\sim0.3$ for Milky Way-like galaxies, along with radio-loud AGNs up to $z>4$ and radio-quiet AGNs up to $z\sim2$. The pilot EMU survey (\cite{Norris2021}) was conducted in the southern field (over $\sim$270 deg$^2$) in the frequency range $800-1088\,$MHz, produced images reaching a sensitivity of $\sim 25-30\, \mu$Jy$\,$beam$^{-1}$, with an angular resolution of $\sim 11-18$ arcsec and resulted in a catalogue of $\sim 220 000$ sources.
At the end of 2022, the main EMU continuum survey started and is planned to reach a better sensitivity and dynamic range with respect to the pilot project.

The next-generation radio astronomy facility SKAO is planned to commence scientific operations by the end of the current decade. One of the primary objectives of SKAO is to investigate the history of cosmic star formation, even reaching extremely distant cosmic epochs.
The initial phase, referred to as SKA1, was originally designed to be executed across two distinct locations, each with different arrangements of baselines. The first site, labelled SKA1-low, comprises 512 stations with 256 antennas each and is located at the Murchison Radio-astronomy Observatory in Western Australia. This array was designed to operate within the frequency range of 50 to 350 MHz. The second site, SKA-mid, was projected to incorporate a combination of 133 SKA 15-meter dishes and 64 MeerKAT 13.5-meter dishes situated at the Karoo site in South Africa. SKA-mid's observation range is planned to range from 350 MHz to 15.4 GHz, with the goal of reaching 24 GHz in the future, and is planned to be divided into at least 4 bands.

The first SKA science data challenge (\cite{Bonaldi2021}) provided the properties of SKA-MID continuum imaging products, addressing the issues associated with their analysis. The images are simulated at three frequencies (150 MHz, 1.4 GHz and 9.2 GHz) and three depths (8 hours, 100 hours and 1000 hours). The rms for the deepest observations reach $\sim 250$, $73$ and 38 nJy in the three bands respectively.

According to projections from \cite{Mancuso2015}, a survey reaching an rms sensitivity of $\sim 0.25 \, \mu {\rm Jy}$ beam$^{-1}$ at 1.4 GHz could potentially identify $\sim 1200$ strongly lensed galaxies per square degree above a $5\sigma$ detection threshold, even at redshifts up to 10. Notably, the impressive angular resolution of SKA1-mid (ranging from 0.03 to 1.4 arcsec) ensures that at least two images from around $30\%$ of these sources will be detected, allowing for the direct confirmation of their lensed nature.

The capabilities of SKA1-mid and its predecessors promise a comprehensive insight into the history of star formation during the EoR. This data will remain unaffected by dust extinction, thereby enabling a thorough exploration of both SFG and radio-quiet AGN populations. 
These observations will achieve unprecedented levels of sensitivity (sub-$\mu$Jy) in the deepest fields. Additionally, they will generate extensive samples over wide areas at the same depth ($\sim \mu$Jy), to date attained only by the smallest and most in-depth radio surveys. 
In conjunction with in-depth multi-wavelength data, this approach will provide an unbiased perspective on the interplay between star formation and nuclear activity throughout cosmic history.

\section{Improving the sub-mm/NIR spectral and continuum information}\label{sec:4}

Aside from the radio band, the launch of JWST in 2021 has inaugurated a new perspective on the distant Universe in the NIR and MIR regime. 
Deep imaging of JWST's imaging cameras (NIRCam and MIRI), demonstrated the ability of JWST to identify optically and NIR dark sources missed even by \textit{Spitzer} observations because of their fainter luminosities. 
By selecting galaxies based on their broadband colours and brightness drop (referred to as the Lyman Break or the Lyman Forest), the exploration of numerous extremely high-redshift (up to $\sim16$) LBGs \textit{candidates} (e.g. \cite{Bradley2022}, \cite{Castellano2022}, \cite{Finkelstein2022}, \cite{Naidu2022}, \cite{Adams2023}, \cite{Atek2023}, \cite{Bouwens2023}, \cite{Donnan2023}, \cite{Finkelstein2023}, \cite{labbe2023}, \cite{Harikane2023}, \cite{Morishita2023}, \cite{Rodighiero2023}, \cite{Yan2023}) was made possible. 
Along with this, the accuracy of JWST's spectrography (performed with NIRSpec and NIRIS) confirmed the presence of galaxies up to $z\sim13.2$ (\cite{Robertson2022}, \cite{Curtis-Lake2023}). 

In addition to these thrilling discoveries, the high-resolution imaging capabilities of JWST to capture the stellar distribution of DSFGs hold a pivotal role in comprehending the mechanisms driving the triggering of starbursts, as well as their subsequent transformation into quiescent galaxies.

In particular, JWST can overcome the high-z DSFGs' extremely low brightness in (rest-frame) optical wavelengths, where the distribution of stars within intense starbursts remains largely unexplored and often beyond the capabilities of the HST and 8–10 m telescopes. 
Only a few galaxies have been detected by IRAC/\textit{Spitzer} in the MIR, but the resolution of these detections is insufficient for studying their structure. JWST, combining sub-arcsecond angular resolution and significantly improved sensitivity compared to previous telescopes across the near- and mid-infrared spectrum (i.e., 1–28 $\mu$m), introduces the opportunity to, for the first time, investigate the stellar light and ionized gas structure within these galaxies at sub-kiloparsec scales. 

NIRCam attains higher angular resolutions and sensitivities compared to HST, more closely resembling those of ALMA. This outcome will allow JWST to explore the spatial segregation between obscured and unobscured star formation, even in HST-dark entities, with an unprecedented level of detail. This factor holds significance in the evaluation of predictions made by evolutionary models. For instance, within the in-situ framework (\cite{Mancuso2016a,Mancuso2016b}, \cite{Mancuso2017}, \cite{Lapi2018}), the stellar component is anticipated to span an area larger than a kiloparsec, while obscured star formation occurs within a notably more concentrated sub-kiloparsec region (\cite{Lapi2018}). This prediction has undergone testing on DSFGs' samples (\cite{Pantoni2021}) and a few strongly lensed systems (e.g., \cite{Massardi2017}), showcasing strong alignment with model projections. The capabilities of JWST will expand these investigations to unprecedented angular resolutions and extend them to fainter and more distant galaxies.

The unmatched angular resolution of MIRI overcomes the challenges of object blending commonly observed in instruments working within comparable wavelength ranges, such as \textit{Spitzer} and WISE.
Furthermore, MIRI imaging holds the capability to delve into the stellar and ionized gas structure in the rest-frame NIR of these galaxies, even during the EoR. For instance, a recent investigation by \cite{Alvarez2023} showcased MIR sub-arcsecond imaging and spectroscopy of the farthest lensed hyper-luminous infrared system discovered to date (SPT0311-58, at z = 6.9). Through observations employing the MIRI IMager (MIRIM) and Medium Resolution Spectrometer (MRS), the study unveiled the structural characteristics of stellar emission (at rest-frame 1.26 $\mu$m) and the ionized medium on kiloparsec scales within the system.

\cite{Kamieneski2023} reported NIRCam high-resolution imaging of an ALMA-selected strongly lensed main-sequence galaxy at z$\sim2.3$. Their analysis reveals significant spatial offsets between UV- and IR-emitting components and suggests the presence of a modestly suppressed star formation activity in the inner kiloparsec, possibly associated with the early stages of the quenching process.

NIRCam, MIRI imaging and NIRSpec spectroscopic observations of the lensing system SPT0418-47 at $z\sim4.2$ from the JWST Early Release Science Program led to the discovery of a DSFG lensed companion (\cite{Peng2023}). The latter features a SFR of $\sim 20\, \rm M_{\odot}\, \rm yr^{-1}$ and near-solar metallicity, highlighting the power of JWST in detecting fainter and previously missed galaxies in the early Universe.

Hence, MIRI and NIRCam possess the capability to distinguish arc-like structures and to discover possible companions within lensed systems at high redshifts, during the peak of Cosmic SFH and beyond. This becomes notably significant when considering intricate lens modelling or conducting SED-fitting analyses, as the resolution and segregation of the background source's emission within the system are important aspects to be taken into account. Additionally, JWST/MIRI can effectively track potential excess originating from obscured AGNs within such sources (\cite{Lyu2023}; \cite{Yang2023}).

Finally, the synergy between JWST and ALMA can be exploited to identify the physical properties of candidate strongly lensed DSFGs and dig into the star formation events. Of particular importance will be the forthcoming ALMA2030 Wideband Sensitivity Upgrade (\cite{Carpenter2023}), which will enhance the efficiency in spectral lines detection, also facilitating the redshift estimation of the lensed sources.



\section{Summary}\label{sec:5}

In this paper we outlined the main surveys and selection methods for identifying strongly lensed DSFGs, emphasizing the importance of gravitational lensing in enabling the study of faint and distant astrophysical sources. We also discussed the observational strategies for studying these lensed systems, both through analyses of statistically significant samples and in-depth studies of individual sources. We highlighted the future perspectives and opportunities offered by upcoming radio band facilities like SKAO and its pathfinders, as well as the new capabilities in the near- and mid-infrared regime provided by JWST.

The advent of SKAO, with its ultradeep radio continuum surveys, promises to enhance our knowledge of the radio properties of DSFGs, distinguishing between star formation and AGN contributions. JWST, with its high-resolution imaging and spectroscopic capabilities in the near- and mid-infrared, opens up new avenues for studying the unobscured stellar component, dust-obscured AGNs and dynamics of DSFGs. Moreover, the synergy with ALMA, particularly given its upcoming sensitivity upgrades, continues to be of great importance for probing the cold dust and molecular gas properties in these galaxies along with their spectroscopic redshift determination.

In conclusion, a comprehensive understanding of high-redshift DSFGs requires a multi-wavelength approach, combining observations from optical/NIR regime to the radio band. Strong gravitational lensing serves as a powerful tool, offering a unique opportunity to explore the properties of these distant galaxies that would otherwise remain inaccessible. With the synergy of ALMA, SKAO, and JWST, we are poised to make significant advancements in our understanding of the processes shaping galaxies in the early Universe and their role in the broader context of galaxy formation and evolution.
\vspace{6pt}

\dataavailability{No new data were created or analyzed in this study. Data sharing is not applicable to this article.} 



\acknowledgments{}

\conflictsofinterest{The authors declare no conflict of interest.}

\appendixtitles{no} 
\appendixstart
\appendix



\begin{adjustwidth}{-\extralength}{0cm}

\reftitle{References}



\bibliography{Giulietti2024}

%


\PublishersNote{}
\end{adjustwidth}
\end{document}